\documentstyle[aas2pp4]{article}

\slugcomment{To be published in {\em The Astrophysical Journal\/}}

\lefthead{Parker et al.}
\righthead{{\em UIT\/} Observations of N\,11 in the LMC}

\newcommand {\UIT} {{\em UIT\/}}

\begin{document}

\title{{\em Ultraviolet Imaging Telescope\/} Observations of OB Stars in the N\,11 Region of the LMC}

\author{Joel Wm. Parker\altaffilmark{1,2},
Jesse K. Hill\altaffilmark{1},
Ralph C. Bohlin\altaffilmark{3},
Robert W. O'Connell\altaffilmark{4},
Susan G. Neff\altaffilmark{5},
Morton S. Roberts\altaffilmark{6},
Andrew M. Smith\altaffilmark{5},
\&
Theodore P. Stecher\altaffilmark{5}}

\altaffiltext{1}{Hughes STX Corporation, Code 681, GSFC/NASA, Greenbelt, MD
20771}
\altaffiltext{2}{current address:  SwRI, Suite 426, 1050 Walnut Street,
Boulder, CO 80302}
\altaffiltext{3}{Space Telescope Science Institute, Baltimore, MD 21218}
\altaffiltext{4}{Astronomy Department, University of Virginia, P.O. Box 3818,
Charlottesville, VA  22903}
\altaffiltext{5}{Laboratory for Astronomy and Solar Physics, Code 681,
GSFC/NASA, Greenbelt, MD 20771}
\altaffiltext{6}{National Radio Astronomy Observatory, Edgemont Road,
Charlottesville, VA  22903}

\begin{abstract}
We present an analysis of far-ultraviolet (FUV: 1300-1800~\AA) and optical
($U$, $B$, and $V$) data of the stellar and nebular content of the OB
associations LH 9, 10, and 13 in the Large Magellanic Cloud region N\,11.  The
FUV images from {\em The Ultraviolet Imaging Telescope\/} strongly select the
hot O and B stars; over 1900 stars were detected in the FUV to a limiting
magnitude of $m_{152} = 17$~mag.  The resulting FUV photometry combined with
optical ground-based data indicate there are approximately 88 confirmed or
candidate O stars in the LH 9, 10, and 13 fields alone (in an area of $\sim
41$~arcmin$^{2}$), and possibly as many as 170 to 240 O-type stars within the
entire 40~arcmin-diameter field of view.
\end{abstract}


\section{Introduction}

N\,11 (Henize 1956) is a spectacular star-formation region in the Large
Magellanic Cloud (LMC); at the center of N\,11 is the OB association
Lucke-Hodge~9 (LH~9; Lucke \& Hodge 1970), and in the north is LH~10, which is
extremely rich in early-type O stars (Parker et al. 1992).  Other OB
associations in the N\,11 region are LH~13 to the east and LH~14 to the
northeast.  The entire region is buried in extensive and complex nebulosity.
Although it resides in the distant northwest outskirts of the LMC, the N\,11
region is the second largest H\,{\sc ii} region in the Magellanic Clouds
according to the H$\alpha$ flux measurements of Kennicutt \& Hodge (1986); only
the starburst region 30~Doradus has a larger H$\alpha$ luminosity.  Other
morphological and evolutionary similarities between N\,11 and 30~Dor have been
discussed by Walborn \& Parker (1992), indicating that N\,11 is a more evolved
(by about 2~Myr) version of 30~Dor.  Analogous to R\,136 in 30~Dor, the dense
star cluster HD~32228 at the core of LH~9 contains a W-R star of type WC 5-6
and evolved O stars.

It is likely that sequential star formation has occurred in N\,11 (Parker et
al. 1992; Rosado et al. 1995), with stellar evolution in LH~9 triggering
formation in the surrounding regions including LH~10.  LH~9 resides in what
appears to be a supernova-evacuated and/or wind-blown bubble and contains no
stars earlier than O6, whereas LH~10 is within a region of strong and highly
variable nebulosity and is rich in very early O stars having at least three and
possibly as many as six O3 stars, as well as half a dozen ``ZAMS O stars''
(Parker et al. 1992).  However, these ground-based studies have not covered the
entire area of the N\,11 region, and it is possible that many O stars remain
undetected.   This is particularly true for the perimeter of the
nebulosity surrounding LH\,9; there have been studies of LH\,10 to the north,
but none of the other regions around LH\,9. The observations made by {\em
Ultraviolet Imaging Telescope\/} (\UIT) provide ideal data for a complete
analysis of the massive star population and evolutionary history of the N\,11
region.

\UIT's UV capabilities and wide field (40 arcmin diameter) with moderate
spatial resolution ($\sim 3$~arcsec) are extremely well suited for massive star
studies in the Magellanic Clouds.  The Clouds are close enough that individual
stars can be resolved, and in the ultraviolet, cool companions of hot stars are
strongly selected against, so binaries are less of a concern.  Magellanic Cloud
OB associations tend to be a few arcminutes in diameter, so
in a single image \UIT\ can obtain UV photometry of cluster members (usually
more than one cluster per image) and background field stars.  This is a great
improvement over the smaller CCDs that sometimes must be mosaiced to obtain
data on a single cluster.  Also, optical photometry is not well suited for
determining temperatures (and therefore, masses) of the hottest, most massive
OB stars, so that follow-up spectroscopy must be obtained to classify stars
before a cluster's initial mass function (IMF) or ionizing flux can be
determined.  On the other hand,  UV photometry can more accurately determine
the temperatures of the hottest stars than can visible photometry, and also
provides data on thousands of stars in the time it takes to get a
classification spectrum of a single star.

In this paper we report first results of the \UIT\ observations of N\,11 and
discuss improved insights these data provide into the stellar content of
this spectacular star-formation region.

\section{Data}

During the Spacelab Astro-2 mission which flew aboard the space shuttle
Endeavour on 1995 March 2-18, \UIT\ obtained more than 700 far-ultraviolet
(FUV) images of nearly 200 celestial targets.  The images include 16
fields in LMC and three fields in the SMC (Parker et al. 1996).  The
\UIT\ observations of N\,11 made on 1995 March 13 consist of three exposures
(39 sec, 197 sec, 986 sec) in the B1 filter, which has a centroid wavelength of
$\lambda = 1521$~\AA, and a bandwidth of $\Delta \lambda = 354$~\AA\ [see
Stecher et al. (1992) for the filter response curve].  The 40 arcmin diameter
photographic images were scanned and digitized with a PDS 1010m
microdensitometer, resulting in images with 1.13 arcsec pixels and point-source
profiles with FWHM$= 3.36 \pm 0.29$ arcsec.  Calibrations were made in the same
manner as for Astro-1 (Stecher et al.  1992), based on laboratory measurements
and data obtained during the missions.  Flux value zeropoints were derived
primarily with comparisons to IUE stars, but also with comparisons to stars
observed by OAO-2, HUT, ANS, GHRS, and other UV instruments (Stecher et al.
1992).  UV magnitudes are defined from these fluxes as: $m = -2.5
\log(F_{\lambda}) - 21.1.$ Astrometry was performed with reference to {\em
HST\/} guide stars (Lasker et al. 1990).  A \UIT\ image of the N\,11 region is
shown in Figure~\ref{fig:n11_glossy}.

\begin{figure*}
\caption[]{The 986 s exposure \UIT\ image of the N\,11 region.
Ground-based fields are indicated by outlines.}
\label{fig:n11_glossy}
\end{figure*}

Stellar photometry on the images was performed with {\sc idl} procedures based
on the {\sc daophot} algorithms (Stetson 1987).  Aperture corrections were
calculated for each image, and small corrections were made in the zeropoint
offsets so that the median difference of the flux for all stars was zero
between the three images (putting all three images on the same zeropoint).  The
final magnitude for each star is the average of its measurements on the three
images weighted by the inverse square of its calculated photometric errors.  A
comparison with IUE observations of three relatively uncrowded stars in the
field show that the UIT and IUE fluxes agree to better than 5\%.

\begin{figure}
\vspace*{2.5in}
\includegraphics{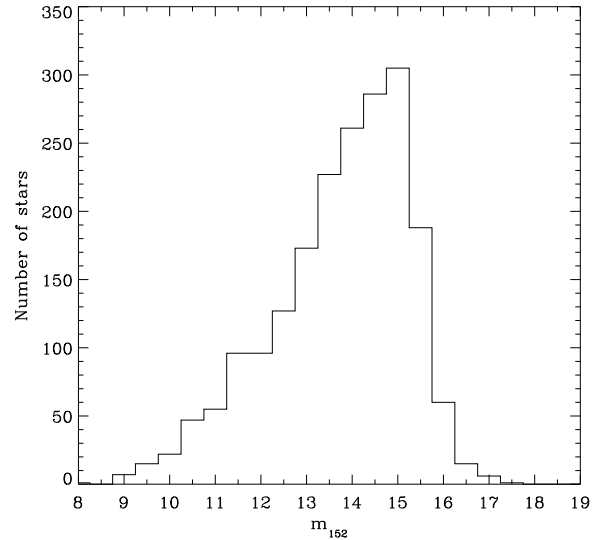}
\caption[]{The observed $m_{152}$ magnitude histogram of the all the stars detected by \UIT\ in the N\,11 region.}
\label{fig:mag_hist}
\end{figure}

Figure~\ref{fig:mag_hist} shows the histogram of observed FUV magnitudes.
The limiting magnitude is $m_{152} \approx 17$~mag, the magnitude of an
unreddened late B-type ($\sim$~B9) star.  The completeness limit, the magnitude
to which we should have detected all stars, is $m_{152} \approx 15$~mag, the
magnitude of an unreddened early B-type ($\sim$~B3) star.  The completeness
limit of the ground-based data of Parker et al. (1992) goes to slightly later
spectral types, so the combined \UIT\ and ground-based dataset is limited by
the FUV data.

The ground-based data are from Parker et al. (1992), roughly $5\times3$~arcmin
fields for each of the LH~9 and LH~10 regions, and from DeGioia-Eastwood,
Meyers, and Jones (1993), covering a $4.1\times2.6$~arcmin region of LH~13.
Outlines of the ground-based fields are shown on the \UIT\ image in
Figure~\ref{fig:n11_glossy}.  These data are from the same source, a 1985
observation run by P. Massey and K.  DeGioia-Eastwood, and were reduced and
calibrated in a similar manner.  In addition to $UBV$ photometry, the LH~9 and
10 data also have spectral types for many of the stars in the field (Parker et
al. 1992).

The cluster HD~32228 at the core of LH~9 has been omitted from this analysis
since the crowded stars therein are unresolved by these observations; cf. the
analysis by Parker et al. (1996) of {\em HST\/} observations of HD~32228 and
other parts of N\,11.  These data will be included in a future paper in which
we will analyze the entire N\,11 region once comparative ground-based data
covering the entire \UIT\ field-of-view have been obtained.

\section{The Color-Magnitude Diagram}

\begin{figure}
\vspace*{2.5in}
\includegraphics{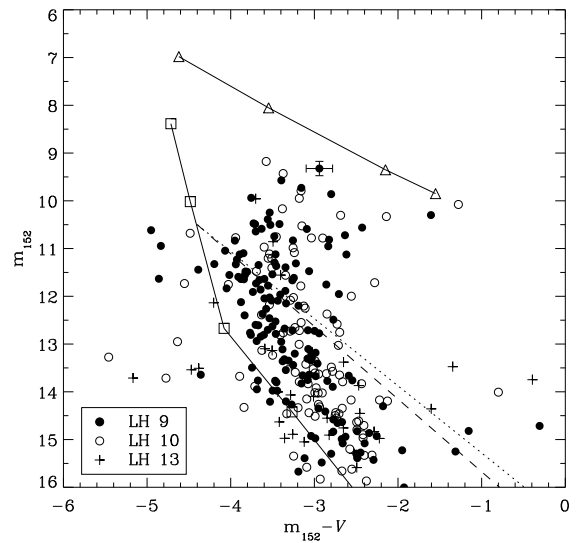}
\caption[]{The observed CMD for LH~9 (filled circles: $\bullet$), LH~10 (open
circles: $\circ$), and LH~13 (crosses: +).  The boxes ($\Box$) connected by a
line show the intrinsic main sequence positions using data from the Fanelli et
al. (1992) UV library, assuming a distance modulus to the LMC of 18.5; the
triangles ($\triangle$) show the location of supergiants.  The dotted line
shows the reddening line for a O9 star assuming the 30~Doradus reddening law,
and the dashed line is for the LMC (non-30~Doradus) reddening law using the
reddening laws from Fitzpatrick (1985).   The error bars shown for one of the
data points indicate a 0.15 mag error in the UV magnitude, and a 0.05 mag error
in the $V$ magnitude.}
\label{fig:obs_cmd}
\end{figure}

Stars were matched between the UV and ground-based data using the closest
positional coincidence within a 3-pixel (3.4~arcsec) tolerance, and the
observed magnitudes were used to create the observed color-magnitude diagram
(CMD) as shown in Figure~\ref{fig:obs_cmd}. A large part of the scatter in the
CMD is a result of the non-uniform extinction in these regions of roughly
$\sigma_{E(B-V)} \sim 0.12$, although intrinsic color variation is probably
also a factor.

Reddenings for LH 9, 10, and 13, as determined by ground-based observations,
are $E(B-V)=0.05$, 0.17, and 0.16 respectively (Parker et al. 1992;
Degioia-Eastwood et al. 1993).  Assuming that the foreground galactic
contribution to the extinction is at least $E(B-V)=0.05$, we can deredden the
observed data by these mean values to obtain the intrinsic CMD, and we find
that  our data are consistent with the previously published average reddening
values.  In Figure~\ref{fig:obs_cmd} we show the reddening lines for the LMC
law and the 30~Doradus law, both from Fitzpatrick (1985).  The difference
between these laws is too small relative to the scatter of the data to make a
definitive determination of the correct reddening law.

\begin{figure}[t]
\vspace*{2.5in}
\includegraphics{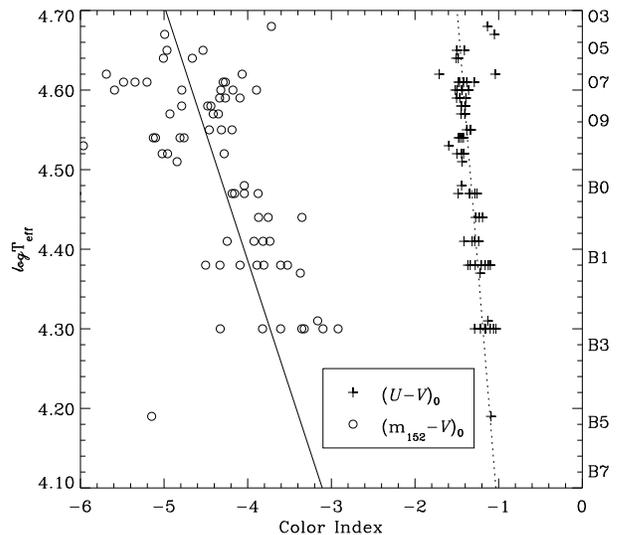}
\caption[]{The relationship between stellar temperature (derived from known
spectral types) and dereddened color for two color indexes: the $U-V$ color
(crosses: +) and $m_{152}-V$ color (open circles: $\circ$).  The linear fit
shows that there is very little dependence of the $U-V$ color (dotted line) on
the temperature for O and at least early B stars, but the $m_{152}-V$ color
(solid line), shows an appreciable color gradient with temperature.}
\label{fig:temp_col}
\end{figure}

The utility of FUV data is shown in Figure~\ref{fig:temp_col}, where the
$(U-V)_0$ and $(m_{152}-V)_0$ color indices are plotted against stellar
temperature.  The stars were dereddened using individual $E(B-V)$ values
calculated by Parker et al. (1992).  Only those stars with spectroscopically
classified spectral types in LH~9 and 10 are shown in the plot, and
temperatures were determined from spectral types, independently of any
photometric information (Parker et al. 1992).  The additional ``leverage'' of
having FUV magnitudes provides a better handle on the temperature (and
therefore, spectral type) of the stars than can be obtained from optical data,
although there is still a large scatter in even the dereddened $m_{152}-V$
colors.  This scatter may be intrinsic but also may be due to the fact that the
FUV colors have been dereddened with $E(B-V)$ data, which is not well-defined
for hot stars because of the degeneracy in $UBV$ colors.  With a FUV extinction
of $R_{\rm FUV}=10.37$, an uncertainty of 0.1~mag in the $E(B-V)$ translates to
an uncertainty of over one magnitude in $E(m_{152}-V)$.  Blending also affects
the scatter, i.e., the star at $\log(T_{\rm eff})=4.19$ in
figure~\ref{fig:temp_col} is a B4~Ia star with a much bluer neighbor that is
resolved in the ground-based data but not in the FUV data, so the star's
$m_{152}-V$ color appears anomalously blue.

\section{The OB Star Population of N\,11}

CMDs allow us to determine the number of O and B stars in the N\,11 region.
The dotted line in Figure~\ref{fig:obs_cmd} indicates the 30~Doradus reddening
line (Fitzpatrick 1985) for an O9 star using data from Fanelli et al.'s (1992)
UV library; the dashed line indicates the LMC law (Fitzpatrick 1985).  All of
the stars above the reddening line are candidate O stars.  A total of 88 stars
appear above the LMC reddening line and all have $V$ magnitudes brighter than
16 mag: 48 stars in LH~9 (a lower limit since this does not include stars in
HD~32228), 34 stars in LH 10, and 6 stars in LH~13.  Only five stars fall
between the 30~Doradus and LMC reddening line, so the results are rather
insensitive to the particular law used.

The number of UV-identified O stars can be compared to the number of O stars
estimated with ground-based photometry.  O stars will have a $Q$-parameter
color of $Q \lesssim -0.85$, where $Q = (U-B) - 0.72(B-V)$, and a visual
magnitude of $V < 16$~mag (the unreddened magnitude of an O9~V star in the LMC
is roughly 14~mag, so this allows for at least 2 magnitudes of extinction or
$E(B-V) < 0.65$, which is larger than the values found in these regions).
Within these limits, 57 candidate O stars are found in the ground-based data:
26 stars in LH~9 (omitting stars in HD~32228), 27 stars in LH 10, and 4 stars
in LH~13.  A total of 43 O-type stars were classified by Parker et al. (1992)
in LH~9 and 10.  No classifications have been made for stars in LH~13.

\begin{figure}
\vspace*{6in}
\includegraphics{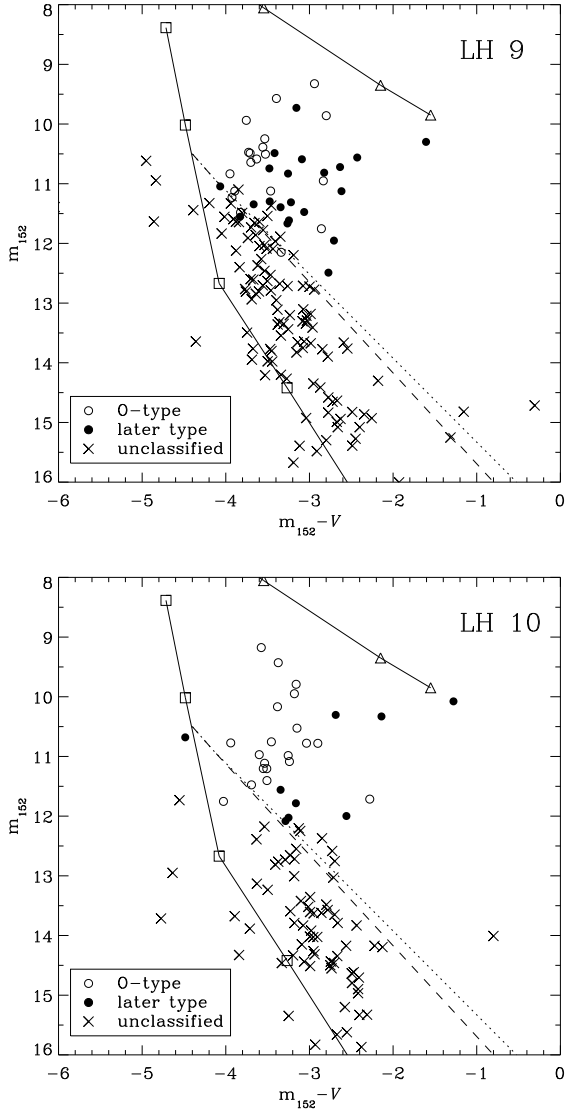}
\caption[]{The color-magnitude diagrams for LH~9 (top) and LH~10 (bottom)
showing the stars with known spectral types, O-type stars (open circles:
$\circ$) and later types (filled circles: $\bullet$), and stars without
spectroscopic classifications ($\times$).  The reddening line and intrinsic
color lines are the same as in Figure~\protect\ref{fig:obs_cmd}.  The majority
of unclassified stars are B-type.}
\label{fig:lh910_cmd}
\end{figure}

Parker et al. (1992) determined IMF slopes for LH~9 and 10, arguing that the
slopes were significantly different, LH~9 being steeper than LH~10, and that
this difference may be the product of the different environments in the
sequential star formation.  However, as discussed above, the FUV data indicate
that the number of O stars in LH~9 may be twice the number estimated from the
ground-based data, whereas only a few more candidate O stars were found in
LH~10 with the FUV data than were found in the ground-based data.  At first one
might think that this could have the effect of significantly flattening the IMF
slope of LH~9 relative to that of LH~10, so that the may be no difference
between the slopes.  However, as shown in Figure~\ref{fig:lh910_cmd}, {\em
all\/} of the unclassified stars are probably late-type O and early-type B
star, since they are positioned at or below the reddening line for an O9 star.
The net effect will be to {\em steepen\/} the IMF slope of LH~9 and increase
the difference with the IMF slope of LH~9.  This strengthens the argument that
LH~9 and 10 may have experienced differing formation triggers and histories,
which could be the result of sequential star formation.  Due to the more
evolved status of LH~9, there are more B stars above the O9 reddening line than
in LH~10.

How many O stars may be in the entire field?  This question can be specifically
answered once we have obtained ground-based data (giving $m_{152}-V$ colors)
for the entire \UIT\ field and followup spectroscopy of a selection of the
stars to verify these results, but we can make a rough estimate here from the
UV magnitudes alone.  As can be seen from the CMDs in the figures shown here,
most O stars have magnitudes brighter than $m_{152} \sim 12.5$, though some
evolved B stars can be as bright as $m_{152} \sim 10.5$.  So, the collection of
``candidate O-type stars'' (COTS) includes not only O-type stars but also stars
that have evolved so that they have temperatures of B-type stars but are
luminous enough to appear at least as bright in the FUV as the faintest O-type
stars.  Approximately 30\% of all the stars brighter than $m_{152} = 12.5$ in
Figure~\ref{fig:obs_cmd} are B stars.  This is consistent with Buscombe's
(1995; Buscombe \& Foster 1994) compilation of stars in clusters with spectral
classifications: of all COTS in LMC clusters (a total of 641 stars) listed in
that catalog, 36\% are luminous (luminosity class I--III) B-type stars.

One might expect the non-cluster field stars would have a larger proportion of
luminous B stars compared to the stars in clusters.  However, the Buscombe
(1995) catalog for field stars in the LMC gives nearly the same fraction, 37\%,
of luminous B stars out of 346 COTS.  Similar results are found in the data
from Massey et al.  (1995), who give spectral types for OB stars in the field
and associations of the Magellanic Clouds.  Massey et al.'s (1995)
``Incompleteness Fields'' in the LMC are a good comparison to the N\,11 region
since these regions include stars in three possible OB associations as well as
stars in the field.  For these regions, out of a total of 43 COTS, 35\% of them
are B stars.  Of all the COTS Massey et al. (1995) classified in the LMC (a
total of 136 stars from their Tables~1 and 5), 41\% of them are B stars.

Another way to estimate the fraction of B stars in a collection of COTS is to
model the distribution assuming an IMF.  A star with a zero-age main-sequence
mass of about 7~${\cal M}_{\odot}$ (an early B V star) is the lowest mass star
that can have a FUV magnitude of $m_{152} < 12.5$ at any point during its
lifetime.  If one assumes that the O and B stars live about 10\% of their
lifetimes in an evolved state that is at low enough $T_{\rm eff}$ to be a B
star and bright enough to have $m_{152} < 12.5$, then for a typical Salpeter
IMF with a slope of $\Gamma=-1.35$, we would find that 25\% of the COTS would
be classified as B-type stars.  Even if one assumes that the IMF slope of the
field stars is as steep as $\Gamma=-4.1$ (Massey et al. 1995), 75\% of the COTS
would be B stars.  However, the addition of young, OB associations in the
region would significantly lower that percentage.

These comparisons of catalogs and modeling of IMFs show it is not unreasonable
to estimate that 30\% to 50\% of the stars in the UIT field brighter than
$m_{152} = 12.5$ might be B stars.  In the entire \UIT\ image, there are 340
stars brighter than $m_{152} = 12.5$, which implies a total O star population
of 170 to 240 stars, four to over five times as many as are currently known in
the region.

Although, more accurate temperatures and masses of these hot stars can be
determined only from time-intensive and methodical classification of spectral
types from spectroscopic observations, the \UIT\ data give us the best possible
photometrically-determined values.  From the \UIT\ data we also can obtain the
most reliable list of candidate O and B-type stars over an unmatched
combination of moderate resolution and large field of view at these
wavelengths.   Even though this preliminary analysis of the N\,11 region is
incomplete (pending current programs to obtain ground-based data covering the
rest of the \UIT\ field and {\em HST\/} observations of the unresolved cluster
cores), we have uncovered a surprisingly large number of previously unknown O
and B-type stars in this one region alone.  This letter provides an exciting
example of how \UIT's rich dataset can be used to obtain unique information on
the hot, massive star population of the Magellanic Clouds and other galaxies.

\newpage

\acknowledgments

We thank M. Fanelli and W. Landsman for useful discussions, J. Offenberg for
his help with the manuscript, and the constructive comments of the referee.
Funding for the \UIT\ project has been through the Spacelab Office at NASA
Headquarters under Project number 440-51. R.~W.~O. was supported in part by
NASA grants NAG 5-700 and NAGW-2596 to the University of Virginia.


\begin{references}
\reference{B1995} Buscombe, W. 1995, ``General Catalogue of MK
	Classifications'' (retrieved from the Astronomical Data Center)
\reference{BF1994} Buscombe, W., \& Foster, B. E. 1994, ``MK Spectral
	Classifications, Eleventh General Catalogue'' (Northwestern University)
\reference {DMJ93} DeGioia-Eastwood, K., Meyers, R. P., \& Jones, D. P. 1993,
	\aj, 106, 1005
\reference{FOBW92} Fanelli, M. N., O'Connell, R. W., Burstein, D., \& Wu, C.-C.
	1992, \apjs, 82, 197
\reference{F85} Fitzpatrick, E. L. 1985, \apj, 299, 219
\reference{H56} Henize, K. G. 1956, \apjs, 2, 315
\reference{KH86} Kennicutt, R. C., \& Hodge, P. W. 1986, \apj, 306, 130
\reference{LSMRJ90} Lasker, B. M., Sturch, C. R., Mclean, B. J., Russell, J.
	L., Jenkner, H. 1990, \aj, 99, 219
\reference{LH70} Lucke, P. B., \& Hodge, P. W. 1970, \aj, 75, 171
\reference{MLDG95} Massey, P., Lang, C. C., DeGioia-Eastwood, K, \& Garmany, C.
	D.  1995, /apj, 438, 188
\reference{PGMW92} Parker, J. Wm., Garmany, C. D., Massey, P., \& Walborn, N.
	R. 1992, \aj, 103, 1205
\reference{Petal96} Parker, J. Wm., et al. 1996, \apj, submitted
\reference{PWSWM96} Parker, J. Wm., Walborn,  N. R., Saha, A., White, R. L. \&
	MacKenty, J. W. 1996, in preparation
\reference{Retal96} Rosado, M., Laval, A., Le Coarer, E., Georgelin, Y. P.,
	Amram, P., Marcelin, M., Goldes, G., \& Gach, J. L. 1996, \aap, in
	press
\reference{S92} Stecher, et al. 1992, \apj, 395, L1
\reference{S87} Stetson, P. B. 1987, \pasp, 99, 191
\reference{WP92} Walborn, N. R., \& Parker, J. Wm. 1992, \apj, 399, L87
\end{references}
\end{document}